# Lindblad's epicycles – valid method or bad science?

Charles Francis[1]

[1] *25 Elphinstone Rd., Hastings, TN34 2EG, UK.*

The study of Galactic orbits in the last eighty years has been dominated by statistical assumptions made because of the lack of empirical evidence available in the early 20th century. Using evidence from Hipparcos and recent radial velocity surveys, Francis and Anderson recently showed that spiral structure is primarily a consequence of gravitational alignments of stellar orbits. I review the mechanism which creates spiral structure, consider the validity of widely held assumptions in galactic dynamics and the implications to notions such as the asymmetric drift and disc heating. I identify a number of fundamental mathematical and physical errors in Lindblad's epicycle theory and in density wave theory. Students should be made aware that these ideas can no longer be considered as science, and authors of textbooks should consider whether they merit anything more than a historical note.

## 1 Background

In popular culture, epicycles have become almost synonymous with bad science; "adding epicycles" refers to a process of introducing fudges to make a theory fit data, when actually the theory needs to be replaced in its entirety. It is generally believed that epicycles were banished from science when Newton solved his equations of motion and showed that it follows from the inverse square law of gravity that planetary orbits are ellipses. So, it comes as something of a surprise to those unfamiliar with galactic dynamics that the galactic orbits of stars are treated in textbooks using a theory of epicycles revitalized by Bertil Lindblad in the 1920s, and used to introduce density wave theory, which, as reinforced by Lin & Shu (1964), by Lin, Yuan and Shu (1969) and by Kalnajs (1973), has been the leading model of spiral structure for nearly 40 years.

It is well known that an epicyclic approximation can be made to any curve. For example, youtube contains an epicyclic approximation to Homer Simpson, http://www.youtube.com/watch?v=QVuU2YCwHjw. It is perfectly possible to approximate any orbit with a system of epicycles; the question is whether doing so has either quantitative or qualitative benefit. Certainly a principle reason for the rejection of Ptolemy's epicycles in the study of planetary orbits is that epicycles do not reflect the underlying dynamics of a system determined by a central force. Nor is there any reason to suppose that epicycles represent the underlying dynamics of galactic orbits.

Galactic orbits are expected to precess because of the matter distribution in the disc and the halo, generating a rosette. Many studies in stellar dynamics rest upon the notion that stars move in near-circular orbits and that their orbits precess at different rates, resulting in a well mixed distribution. E.g. the calculation of the LSR from Strömberg's asymmetric drift equation (e.g. Dehnen & Binney, 1998). However, there have been a number of recent studies challenging the assumption of a well mixed distribution (Skuljan, Hearnshaw & Cottrell, 1999; Fux 2001; Dehnen, 1998; Famaey et al., 2005, Chakrabarty, 2004 & 2007, Quillen 2003, Minchev & Quillen 2006, de Simone et al, 2004, Chakrabarty & Sideris, 2008). In a recent study, Francis and Anderson (2009b; herein FA09b) described how mutual gravity between stars causes orbits to align on approximately logarithmic spirals, and showed from observations of the galactic gas distribution and the motions of over 20 000 stars in the solar neighbourhood that the Milky Way is a two-armed spiral. Contrary to conventional wisdom, stars do not move through the arms on near-circular orbits, but move along an arm on the inward part of their orbits, leave the arm soon after pericentre, cross the other arm on the outward part and rejoining the original arm before apocentre (a brief review is given in section 2). The distribution does not become well mixed over the course of time. Instead orbital rosettes become aligned with spiral arms. As a result, a number of conventional calculations fail (section 2.8).

Epicycles have only been used to study orbits in galactic dynamics. In other fields, perturbations to an orbit due to alteration from the potential due to a central mass are studied using the eccentricity vector or the Laplace-Runge-Lenz vector (see e.g. Arnold, 1989; Goldstein, 1980). It is appropriate, therefore, to assess whether Lindblad's epicycles have any value in the study of galactic dynamics. Section 3 will describe Lindblad's epicycles, and show that the complexity of the treatment has no benefit over a more conventional analysis, but has rather obscured dynamics and lead to mistakes in analysis.

## 2 Gravitationally Aligned Rosettes

### 2.1 The Eccentricity Vector

For an elliptical orbit the eccentricity vector is defined as the vector pointing toward pericentre and with magnitude equal to the orbit's scalar eccentricity. It is given by

$$\boldsymbol{e} = \frac{|\boldsymbol{v}|^2 \boldsymbol{r}}{\mu} - \frac{(\boldsymbol{r} \cdot \boldsymbol{v})\boldsymbol{v}}{\mu} - \frac{\boldsymbol{r}}{|\boldsymbol{r}|}, \qquad (2.1.1)$$

where $\boldsymbol{v}$ is the velocity vector, $\boldsymbol{r}$ is the radial vector, and $\mu = GM$ is the standard gravitational parameter for an orbit about a mass $M$. For a Keplerian orbit the eccentricity vector is a constant of the motion. Stellar orbits are not strictly elliptical, but the orbit will approximate an ellipse at each part of its motion, and the eccentricity vector remains a useful measure (the Laplace-Runge-Lenz vector, which is the same up to a multiplicative factor, is also used to describe perturbations to elliptical orbits).

## 2.2 Orbital Precession

In the solar system, mass is concentrated at the Sun, and planetary orbits are ellipses. In a spiral galaxy, mass is distributed throughout the disc and in the halo. As a star moves towards pericentre, the gravitational mass drawing it towards the galactic centre is less than it would be if all the mass of the galaxy was concentrated at one central point. As a result, the orbit of the star is less curved at pericentre, and more curved at apocentre, than an ellipse, and the orbit precesses, forming a rosette (figure 1). However, if we imagine looking at the motion from above, from a platform rotating at the rate of precession; the orbit will be approximately elliptical with the galactic centre at the focus.

## 2.3 Spiral Structure

An equiangular spiral structure can be constructed by enlarging an ellipse by a constant factor, $k$, centred at the focus and rotating it by a constant angle, $\tau$, with each enlargement (figure 2). The pitch angle of the spiral depends only on $k$ and $\tau$, not on the eccentricity of the ellipse. The pitch angle of a given spiral galaxy is directly related to the orbital eccentricities of stars in that galaxy. For a given pitch angle, ellipses with a range of eccentricities can be fitted to the spiral, depending on how narrow one wants to make the spiral structure and what proportion of the circumference of the ellipse one wants to lie within it. Higher eccentricity orbits fit spirals with higher pitch angles. Thus we can understand the spiral structure of the arms, provided that a mechanism can be exhibited which causes orbits to align with the arm, and which ensures that the angular rate of orbital precession is independent of orbital radius.

## 2.4 Spiral Potential

The gravitational potential of a spiral galaxy can be described as a spiral-grooved funnel (figure 3). The grooves represent the gravitational field of the spiral arms. A star near apocentre, the slowest part of its orbit, will tend to fall into a groove and then follow the groove, picking up momentum as it goes. Eventually, the star gains enough momentum to jump free of its groove. It crosses over the next-highest groove, then falls back to a higher point in its original groove (an animation is shown at http://rqgravity.net/SpiralStructure). At the same time, the funnel rotates slowly backwards due to orbital precession.

As stars are drawn into an arm, the gravitational field of the arm grows stronger, drawing greater number of stars into the arm. This mutual gravity between stars reinforces spiral structure, and the potential field of the arms locks the rate of orbital precession to spiral pattern speed for a wide range of orbits. FA09b describes the evolution from flocculent through multiarmed to grand-design bisymmetric spirals.

## 2.5 Star Formation in Spiral Arms

Under gravity, gas clouds follow similar orbits to stars (figure 4). Gas in the arm is in turbulent motion, as gas clouds seek to cross in the arm and gain velocity as they approach pericentre. Whereas stars rarely collide because of their small size compared to space between them, when outgoing gas from one arm meets ingoing gas in another arm, collisions between gas clouds create regions of higher pressure, and greater turbulence. Pockets of extreme pressure due to turbulence generate the molecular clouds in which new stars form.

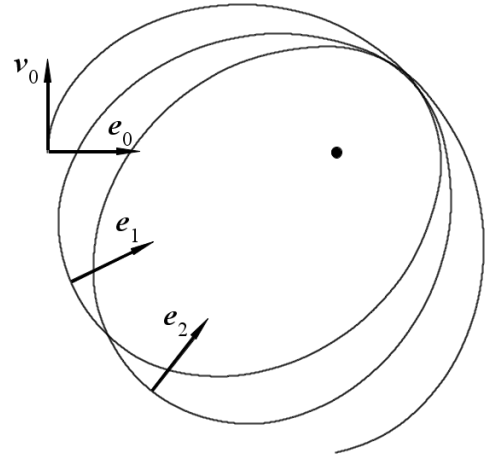

**Figure 1:** The eccentricity vector of an orbit regresses for a central core plus disc. Regression has been exaggerated by increasing the mass of the disc relative to the core (by comparison with the Milky Way). The simulation used a central mass of 35 billion solar masses, a disc density $0.3e^{-R/3}$ billions solar masses per kpc$^2$, initial radius 8kpc and initial velocity 190kms$^{-1}$.

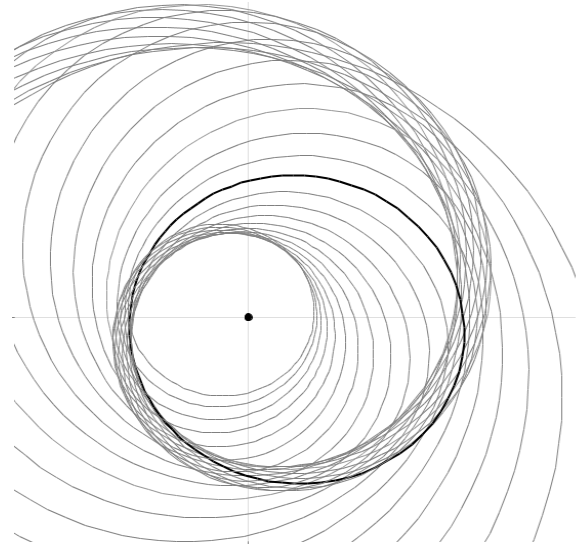

**Figure 2:** An equiangular spiral with a pitch angle of 11°, constructed by repeatedly enlarging an ellipse with eccentricity 0.3 by a factor 1.05 and rotating it through 15° with each enlargement. Lower eccentricity ellipses produce a narrower structure. Ellipses with eccentricity greater than about 0.25 have more than half their circumference within the spiral region. Ellipses with eccentricity greater than about 0.35 produce probably too broad a structure to model a spiral arm with this pitch angle, but give a good fit for spirals with greater pitch angle.

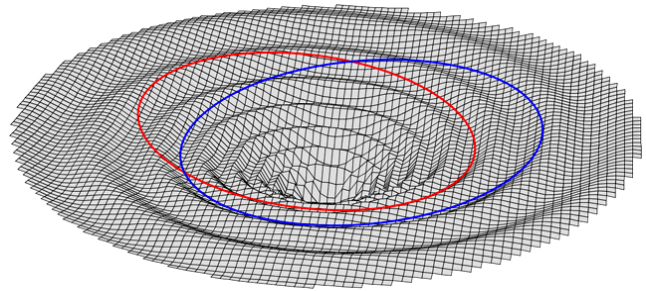

**Figure 3:** The gravitational potential of a bisymmetric spiral galaxy, showing the alignment of elliptical orbits with troughs in the potential.

*2.6    Formation of Bisymmetric Spirals*

In a multi-arm spiral, outgoing gas meeting an arm would have greater mass than ingoing gas in the arm. This would tend to remove gas from the arm. In a two armed spiral, the gas in the arm has greater mass. Thus, a two-armed gaseous spiral can be stable, whereas multi-armed gaseous spirals cannot. Outgoing gas applies pressure to the trailing edge of a spiral arm. If one gaseous arm advances compared to the bisymmetric position, the pressure due to gas from the other arm will be reduced. At the same time, pressure on the retarded arm due to outgoing gas from the advanced arm will be increased. Thus gas motions preserve the symmetry of two-armed spirals.

*2.7    Spiral Structure of the Milky Way*

Pitch angles of around 10-15° (e.g. figure 2), as used in four armed spirals (e.g. Georgelin & Georgelin, 1976; Russeil, 2003; Levine, Blitz and Heiles), correspond to orbital eccentricities in a range greater than about 0.25, and are incompatible with the eccentricity distribution of the local stellar population (the eccentricity distribution in figure 5 is based on the LSR calculated by Francis and Anderson (2009a; herein FA09a), but the incompatibility remains for any reasonable value of the LSR). A two-armed spiral with pitch angle around 5° gives a good fit to the bulk of observed eccentricities in a range from about 0.1 to about 0.2, and also gives a good fit with the observed HI distribution (FA09b).

A two-armed spiral necessitates a little care to avoid confusion in naming the arms, because traditionally named sectors with the same name lie on different arms (figure 6). Orion is not a separate spur, but is a part of a major arm connecting Perseus in the direction of rotation to Sagittarius in the direction of anti-rotation. We have called this major spiral arm the Orion Arm. The Orion arm contains Norma, Perseus, Orion, Sagittarius, and Cygnus sectors. The Centaurus arm contains Sagittarius, Scutum-Crux, Cygnus, and Perseus sectors. The solar orbit is shown in approximation, together with its major axis and latus rectum.

*2.8    The Structure of the Local Velocity Distribution*

The velocity distribution (figure 7) contains motions of new stars (Pleiades stream), a flow of stars in the spiral arm on the inward part of their orbits, stars on the outward part of their orbit crossing the Orion arm (Hyades stream), moderately young stars which have not achieved typical orbits in the arm (Sirius stream) and older stars in more eccentric orbits, (Hercules and Alpha Ceti stream). Thus the distribution is highly structured, and traditional measures such as dispersion and mean velocity have little meaning for the population as a whole. As a result, it is necessary to put aside typical text book analyses.

For example, the Oort constants have no meaning, the LSR cannot be calculated by estimating the asymmetric drift, and there is no evidence for disc heating — the hypothetical process by which scattering events are supposed to cause the random velocities of stars to increase with age (Jenkins, 1992). There is an increase in velocity dispersion with colour, up to Parenago's discontinuity (Parenago, 1950). Dehnen & Binney (1998) suggest that the reason for Parenago's discontinuity is the heating of the disc. However, disc heating appears to contradict the second law of thermodynamics and relies on the mistaken belief that velocity dispersion is a measure of randomness. The truth is that dispersion rises with age because spiral structure supports stable orbits with greater than normal eccentricity. Old stars can be found in these orbits, while young stars rarely enter them (FA09b, section 10).

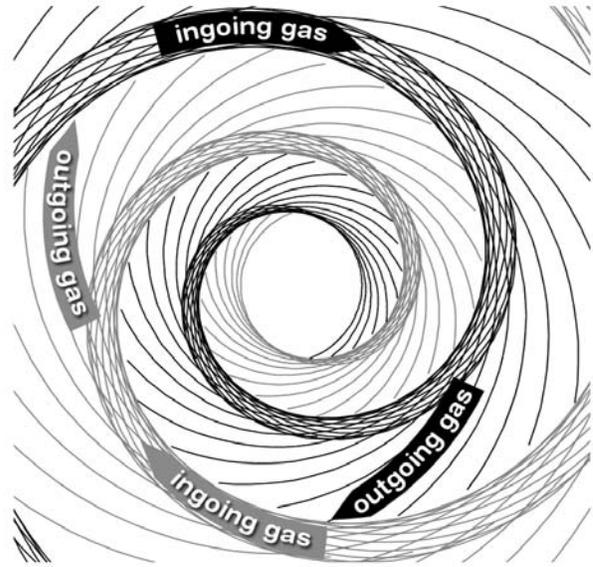

**Figure 4:** Gas motions in a bisymmetric spiral galaxy.

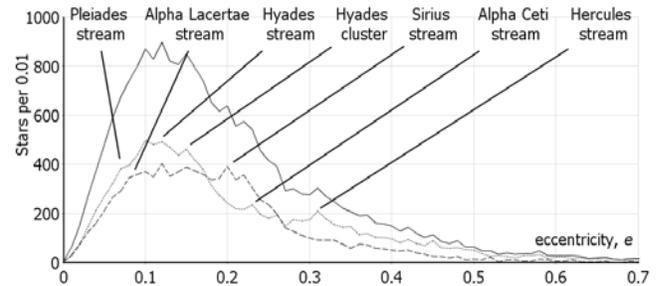

**Figure 5:** Eccentricity distribution (from FA09a) for stars closer to apocentre (dotted) and stars closer to pericentre (dashed), as defined by position with respect to the semi-latus rectum.

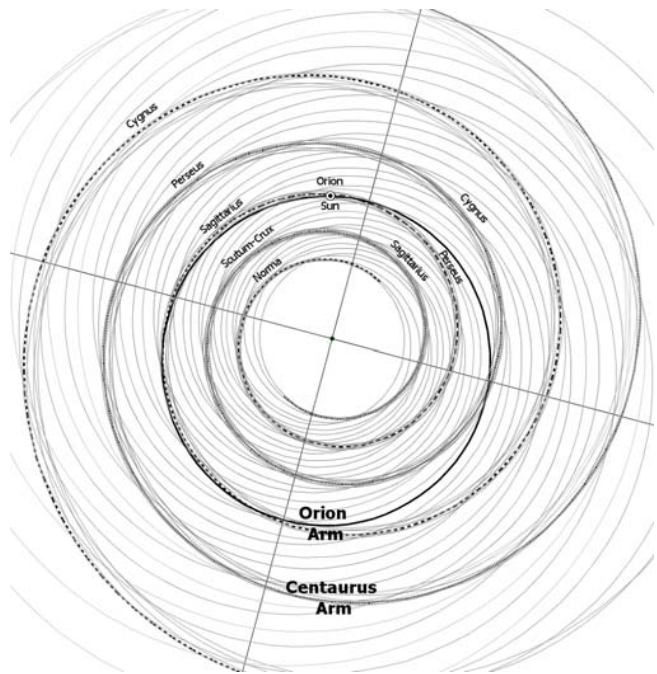

**Figure 6:** Two-armed spiral model of the Milky Way, based an angular increment of $\tau = 30°$ for each 105% enlargement, giving a pitch angle of 5.44°. The solar eccentricity, 0.138, has been used for the diagram.

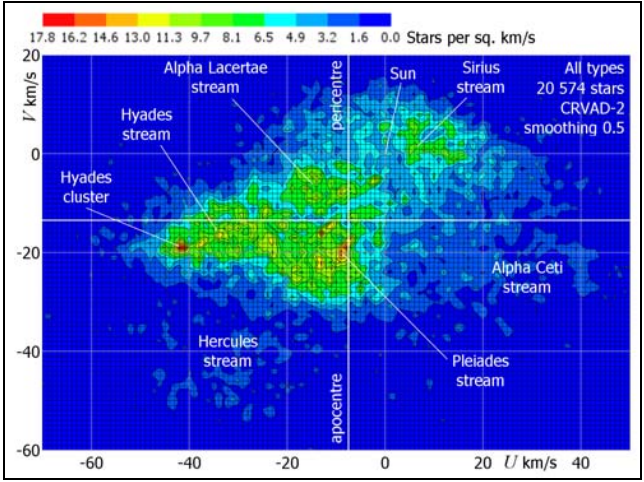

**Figure 7:** The distribution of *U*- and *V*-velocities from FA09a, showing the Hyades, Pleiades, Sirius, Alpha Ceti, Alpha Lacertae and Hercules streams.

## 3 Errors in Density Wave Theory

### 3.1 The Winding Problem

The analysis of spiral structure is often introduced with a description of the winding problem. Lindblad argued that the arms cannot consist of stars orbiting together in approximately circular orbits, because orbital velocity decreases with orbital radius, so that a spiral consisting of regions of greater density would wind up over a period of time. Unfortunately, by considering the orbital velocity of stars, this analysis has misdirected later investigations. Stars are not bound in any form of solid structure. Nor is there any prior reason to assume that their orbits can be treated as nearly circular. The evidence of near-circular orbits for moons and planets in the solar systems suggests that tidal forces lead to a decrease in eccentricity over large numbers of orbits, but typical stars at the solar radius will have orbited the Milky way less than about 100 times — far too few to expect the system to be close to a circular equilibrium.

Spiral structure as described in section 2 is in fact independent of orbital velocity; the winding problem as described by Lindblad does not apply. Stars do not individually come together and share similar orbits as would be suggested by a solid motion ("material arms"). Rather stellar paths align on the spiral arm for much of the orbit. The behaviour of the spiral over time is then determined by the rate of precession of the orbits, not by orbital velocity. The gravitational potential of the spiral modifies the precession of the orbits, so as to preserve and reinforce spiral structure, and to ensure that they all regress at the same rate, the rate of spiral pattern speed.

### 3.2 The Epicyclic Approximation

Lindblad's epicycles perturb a circular orbit by superimposing an elliptical motion. As is seen in figure 8, for low eccentricities it is possible to approximate a precessing orbit. However, this is not useful. After the effort, and with the inaccuracy, of introducing the epicyclic approximation, no more has been said than that the orbit is precessing oval, or a rosette. To take the analysis any further, and calculate, for example, the rate of precession, would require further corrections because the epicyclic approximation is only good for low eccentricities, and would require knowledge of the gravitational potential throughout the orbit. Epicycles cannot be used as an iterative method, because there is no practicable way to continue the iteration. If fluctuations in potential due to spiral arms, the bar, and satellites, are to be considered, then analysis is only possible by numerical methods. However, if a computer will be used to provide a numeric solution, there is no point in starting with an approximation. It would be better, and simpler, to find a numerical solution directly from the equations of Newtonian gravity.

### 3.3 Closed Orbits

A closed orbit is one which returns to apocentre at the same point after a number of cycles, and then repeats the same path. We can close the orbit in figure 8 by using coordinates rotating at just such a rate that apocentre moves so as to coincide with a previous apocentre after a number of orbits. We can calculate an equation for the rate of rotation of the coordinate system as follows:

Angular speed, or angular frequency, of circular motion: $\Omega$.
Angular speed of radial oscillation (i.e. elliptical motion): $\kappa$.
Period of radial oscillation: $2\pi/\kappa$.
Angular distance, $\widehat{AB}$, after $m > 0$ radial oscillations:

$$\widehat{AB} = 2\pi n \pm \Omega \frac{2\pi m}{\kappa}, \ n \in \mathbb{Z}. \tag{3.3.1}$$

**Mistake 1:** Standard treatments (e.g. Binney & Tremaine, 1987; Binney & Merrifield 1998; Carroll & Ostlie, 1996) overlook the possibility of a minus sign in (3.3.1). The orbit may be closed by rotating through either the major or the minor arc $\widehat{AB}$.

Angular speed of rotating reference frame: $f$.

To establish a closed orbit in the rotating coordinates, after $m$ periods of the radial oscillation, we require,

$$f \frac{2\pi m}{\kappa} = \widehat{AB} = 2\pi n \pm \Omega \frac{2\pi m}{\kappa}, \tag{3.3.2}$$

$$f = \frac{n}{m}\kappa \pm \Omega. \tag{3.3.3}$$

The usual solution uses plus and $m = -2n$.

$$f = \Omega - \frac{1}{2}\kappa. \tag{3.3.4}$$

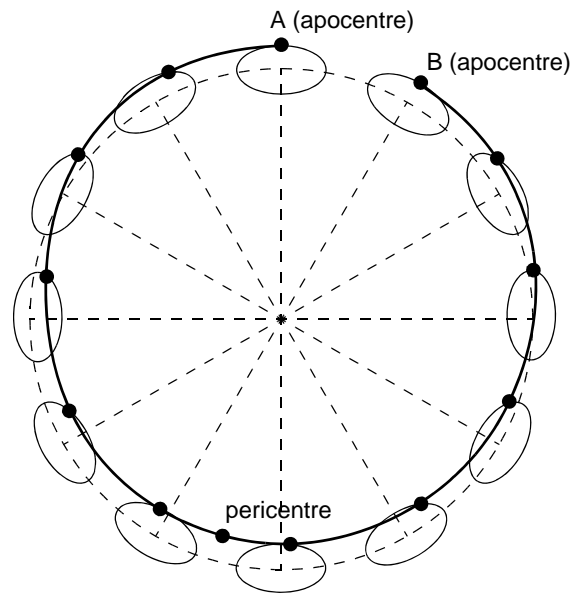

**Figure 8:** Lindblad's epicycles approximate a precessing ellipse by superimposing an elliptical motion on a circular orbit.

**Mistake 2:** This ignores imaging issues with $m > 1$, such that the rotating coordinate system shows more than one orbit.

**Mistake 3:** The use of plus in (3.3.3) means the rotation of coordinates is in the direction of orbital motion, but actually orbits regress.

The correct rate for spiral pattern speed requires minus and $m = n = 1$. Hence

$$f = \kappa - \Omega .\qquad(3.3.5)$$

*3.4   Density Wave Theory*

Spiral structure is usually explained using the density wave theory of Lin and Shu. (1964). In this model it is said that stars move through the arms on near-circular orbits, and the arms consist of dense regions analogous to regions of heavy traffic on a motorway.

**Mistake 4:** A simple analogy with patches of heavy traffic fails because a wave effect would require that stars slow down when they approach a dense region, but the gravity of the dense region would cause them to speed up. Even if the theory were right, the analogy would be wrong.

**Mistake 5:** The claim that stars move through the arms on near-circular orbits is empirically incorrect (FA09a)

Following Kalnajs (1973), density wave theory is usually explained by means of a diagram (figure 9) constructed by enlarging and rotating ovals. The orbits are an epicyclic approximation in coordinates rotating at a rate $\Omega - \kappa/2$ and appear as ovals centred at the galactic centre.

**Mistake 6:** (following from mistake 2): Even if the rotation of the ovals were correct (it is not), because of the rate of rotation of coordinates in which it is drawn, figure 9 would show a single spiral twice, not a bisymmetric spiral.

**Mistake 7:** (to be replaced by mistake 8): The increase in density shown as spirals in the figure represents only a small proportion of the orbit. If stars were placed randomly, one on each oval, this would lead to a very small increase in stellar density on the "arms", an order of magnitude less than is observed.

**Mistake 8:** ("correcting" mistake 7): The orbits in the figure are imagined as gas in lamina flow. Then the increase in gas density in the spiral pattern is assumed to instigate star creation in the arms. But this increase in density would not be sufficient to initiate star formation, and spiral arms contain a substantial increase in density of old as well as new stars.

**Mistake 9:** (following from mistakes 6 & 8): If the ovals were to represent gas motion, it could not be treated as lamina flow, because while paths do not cross in rotating coordinates, they do cross in physical space (figure 10). Taken as the motions of gases, figure 9 is impossible.

**4   Conclusion**

Understanding the cause of spiral structure has appeared a difficult problem for about eighty years, ever since it was clearly recognised that spiral nebulae are other galaxies. It is a many-body problem with unknown initial conditions, and its solution involves the turbulent motion of interstellar gas in addition to the motions of the stars. Nonetheless, once known, the solution is remarkably straightforward, and is confirmed by the empirical evidence of the local velocity distribution and the neutral hydrogen distribution.

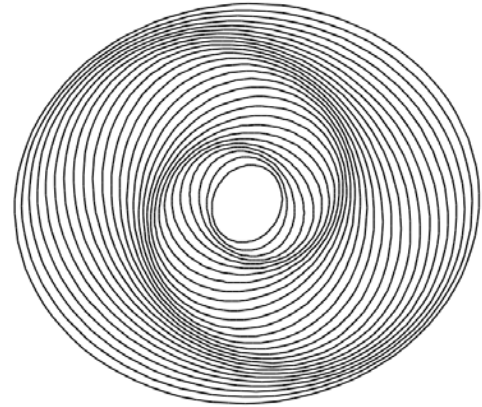

**Figure 9:** Enlarging and rotating ovals aligned at the centre generates a two-armed spiral.

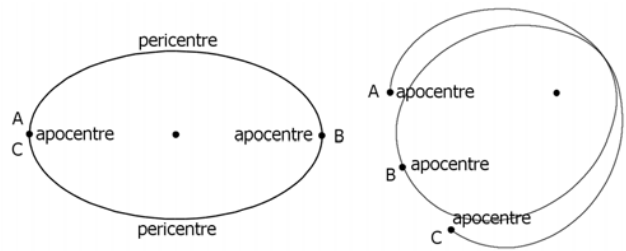

**Figure 10:** (left) Closed orbit in rotating coordinates with $f = \Omega - \kappa/2$, and (right) the corresponding path in non-rotating coordinates.

The introduction of a working model for spiral galaxies will lead to important changes in the study of galactic dynamics. Many of the treatments found in textbooks are seen not to apply. Perhaps of greater concern to astrophysics as a science is that some of those treatments are not even internally consistent. When mathematicians have addressed the problem of orbits in a gravitational field differing from that of a central mass distribution, they have perturbed the eccentricity vector or the Laplace-Runge-Lenz vector. Lindblad's epicycles have distracted astrophysicists from this type of analysis and substituted a model which is conceptually more complex, and which, because numerical solution is still required, adds nothing to what is already known of galactic orbits. The analysis has been based empirical assumptions about orbits which were not justified from data at the time, and which have since proven false, and it contains elementary mathematical mistakes which have been compounded by further mistakes in the development of density wave theory.

The implication to astrophysics is severe. The motions of stars are governed by known mathematical laws. Astrophysics is, or at least it should be, a mathematical science. One should therefore expect that theories in astrophysics are subjected to rigorous mathematical scrutiny. Regrettably, the degree of scrutiny applied to Lindblad's epicycles and to density wave theory has been seriously lacking. Students should be made aware that these ideas can no longer be considered as science, and authors of textbooks should consider whether they merit anything more than a historical note.